# Geometric Superconducting Diode Effect in an NbN Nanoring

**Tianyu Li[1,2], Peiyuan Huang[1,2], Jiong Li[3], Jiyao Shang[1,2], Nuo-Zhou Yang[1,2], Wuyue Xu[1,2], Wen-Cheng Yue[1,2,4,*], Yang-Yang Lyu[1,2,*], Chong Li[1,5,*], Yihuang Xiong[1], Xuecou Tu[1,2], Tao Tao[4], Xiaoqing Jia[1,2], Qing-Hu Chen[3], Huabing Wang[1,2], Peiheng Wu[1,2], and Yong-Lei Wang[1,2,4,*]**

[1]Research Institute of Superconductor Electronics, Nanjing University, Nanjing 210023, China
[2]Purple Mountain Laboratories, Nanjing 211111, China
[3]School of Physics, Zhejiang University, Hangzhou 310027, China
[4]State Key Laboratory of Spintronics Devices and Technologies, Nanjing University, Suzhou, China
[5]School of Microelectronics, Nanjing University of Science and Technology, Nanjing, Jiangsu 210094, China

E-mail: wenchengyue@nju.edu.cn; yylyu@nju.edu.cn; chongli@njust.edu.cn; yongleiwang@nju.edu.cn

## Abstract

Superconducting diodes, which exhibit nonreciprocal critical currents, are promising building blocks for low-power cryogenic electronics and superconducting circuits. Existing superconducting diode platforms commonly rely on Josephson junctions, multilayer heterostructures, ferromagnetic elements, gate-difined structures. Here, we demonstrate a geometrically induced superconducting diode effect realized in a structurally minimal, single-materials NbN nanoring, where inversion-symmetry breaking is introduced solely by the asymmetric geometry. The device exhibits pronounced and polarity-switchable critical-current nonreciprocity. Systematic magnetic-field and temperature-dependent measurements reveal that, at low fields, the applied magnetic field redistributes the critical current asymmetrically between opposite bias directions without significantly reducing the overall superconducting current-carrying capability. Moreover, the maximal nonreciprocity and diode efficiency exhibit distinct temperature dependence: the maximal diode efficiency follows the evolution of the energy gap, whereas the maximal nonreciprocity is more closely associated with the superfluid density. These results establish asymmetric superconducting nanorings as a minimal geometric platform for studying nonreciprocal superconducting transport and provide a simple design principle for future superconducting electronics.

## 1. Introduction

The superconducting diode effect—nonreciprocal superconducting transport without energy dissipation—has emerged as a key enabling mechanism for next-generation superconducting electronics, low-power cryogenic logic, and quantum information hardware. Fundamentally, achieving a diode effect requires the simultaneous breaking of time-reversal symmetry and spatial inversion symmetry, resulting in asymmetric critical currents for positive and negative bias directions[1-15]. Time-reversal symmetry is typically broken by applying an external magnetic field[7-11, 15-18], incorporating ferromagnetic layers or magnets[5-7, 19-21], or adopting materials with intrinsically time-reversal-symmetry-breaking superconducting states[2, 22]. Spatial inversion symmetry is broken geometrically—through asymmetric structures[8-10, 19, 21], gate-controlled potentials[23-27], magnetic nanostructures[7, 28, 29], or engineered defects[4, 30]. The coexistence of these two symmetry-breaking mechanisms results in direction-dependent superconducting trajectories and unequal critical currents $I_{c+}$ and $I_{c-}$, producing diode behavior[31]. Despite rapid progress, many superconducting diode architectures rely on complex multilayers[9, 32-34], hybrid materials[19, 20], or Josephson junctions[1, 17, 18, 35-37]. Such complexity increases fabrication difficulty and limits scalability, often imposing constraints in reproducibility and circuit-level integration. Therefore, it remains important to establish a simpler geometric route in which superconducting diode behavior is generated directly by the shape of a single superconducting nanostructure.

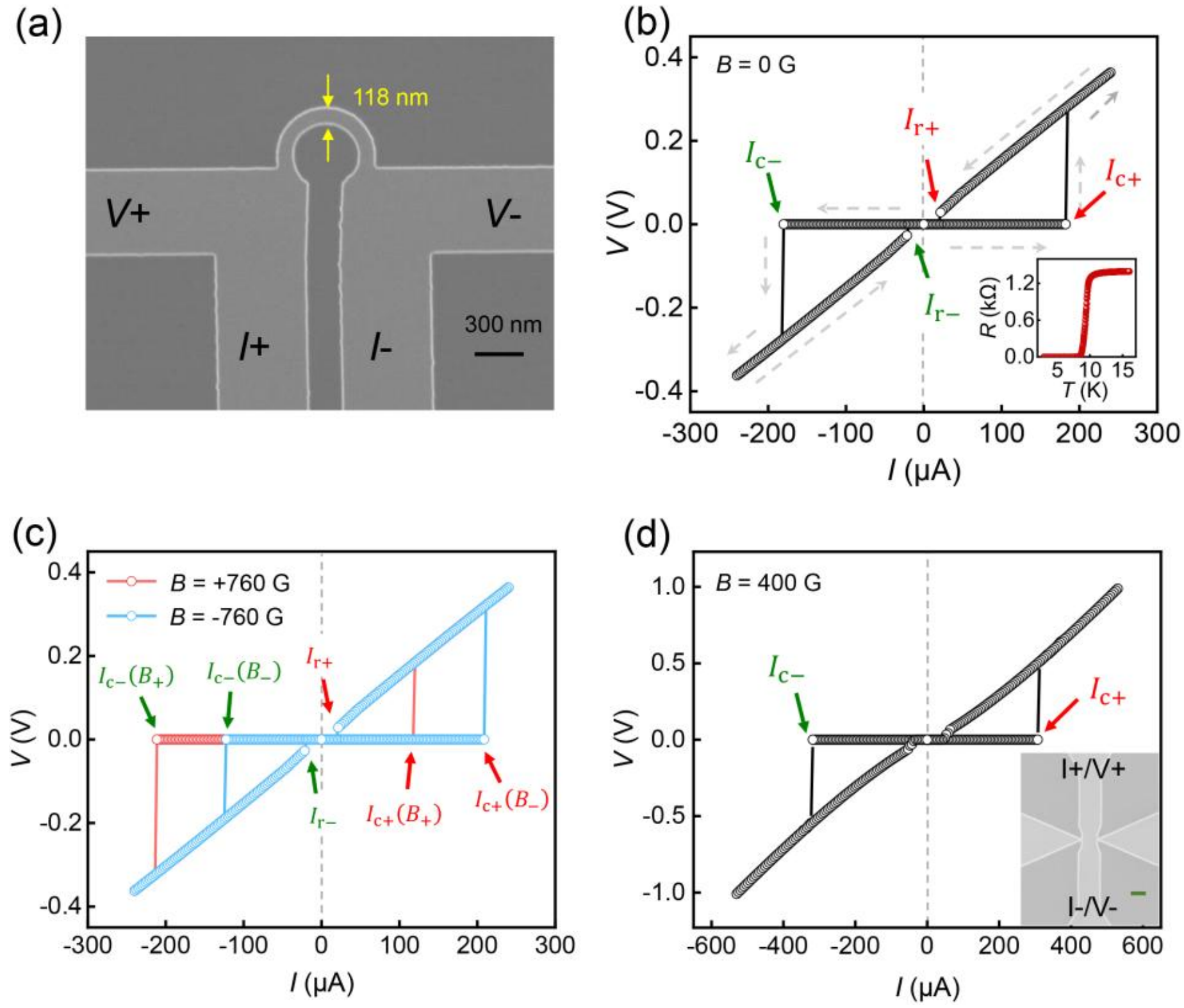

**Figure 1. Superconducting diode effect in an NbN nanoring.** (a) SEM image of the nanoring superconducting diode device. (b) *I-V* characteristics measured at zero magnetic field. The curve exhibits symmetric behavior with negligible nonreciprocity. Inset: Temperature-dependent resistance of the superconducting nanoring, showing a critical temperature $T_c = 8.6$ K. (c) *I-V* characteristics measured at $T = 2$ K under the magnetic fields of +760 G (red) and -760 G (blue), revealing a significant and switchable nonreciprocitiy in critical currents. (d) *I-V* characteristics of a symmetric device measured as a reference. Inset: SEM image of the symmetric device. Scale bar = 500 nm.

Here, we demonstrate a structurally minimal superconducting diode formed by an asymmetric NbN nanoring. Distinct from previous superconducting diode platforms based on Josephson junctions, multilayers, hybrid structures, ferromagnetic elements, gate-controlled potentials, our device uses only the asymmetric annular geometry of a single-material NbN nanowire to break spatial inversion symmetry. The device features a compact nanoscale geometry that inherently breaks spatial inversion symmetry and supports pronounced critical-current nonreciprocity. Despite its minimal structure, the diode exhibits strong rectification and clear superconducting–normal switching behavior, highlighting the effectiveness of geometric design in enabling diode functionality without requiring complex material stacks or engineered magnetic components. These results establish curvature-induced edge asymmetry as a minimal geometric design principle for realizing nonreciprocal superconducting transport in a single-material superconductor.

## 2. Results and discussion

Our superconducting diode device of an asymmetric superconducting nanoring [Fig. 1(a)] was fabricated using a single-layer niobium nitride (NbN) thin film with a thinness of 10 nm. The nanoring has an inner diameter of 600 nm and a linewidth of approximately 118 nm. Transport experiments were conducted employing a standard four-probe method, with the magnetic field applied perpendicular to the sample plane. The fabrication and measurement details are described in the Methods section, with additional process parameters provided in the Supporting Information. Thdevice exhibits a superconducting transition temperature of $T_c = 8.6$ K [inset of Fig. 1(b)].

Current–voltage characteristics were first measured under zero external magnetic field, where the *I-V* curves remain fully symmetric with equal positive and negative critical currents [Fig. 1(b)]. This demonstrates that geometric asymmetry alone does not generate a measurable diode effect under zero field. When an out-of-plane magnetic field $B$ is applied, the device exhibits pronounced critical-current nonreciprocity [Fig. 1(c)]. For example, at $B = +760$ G [Fig. 1(c)], the extracted critical currents under reversed currents are $I_{c+} = 117$ μA and $I_{c-} = -210$ μA, giving rise to a non-reciprocal component $\Delta I_c = 93$ μA . When a current $I$ in the range of $|I_{c+}| < I < |I_{c-}|$ is applied, the device switches to the normal-resistance state in the positive current direction while remaining in the superconducting state (complete zero-resistance state) in the negative direction. This leads to a one-way supercurrent transport, demonstrating a superconducting diode effect. Reversing the magnetic field reverses the polarity of the nonreciprocity [Fig. 1(d)], confirming a magnetic-field-switchable diode behavior.

Edge-controlled vortex motion and flux-flow instability have been quantitatively studied in narrow superconducting structures, for example in direct-write Nb-C superconductors, where a stable flux-flow response enables analysis of ultrafast vortex motion[40]. In the present NbN nanoring, however, the enlarged $I$–$V$ curve shown in Fig. S1 does not reveal a clearly resolved intermediate flux-flow branch. Instead, the device switches abruptly from the superconducting state to the normal-resistance state. Therefore, a direct quantitative flux-flow analysis, such as extracting vortex velocity from a continuous flux-flow voltage, is not applicable here. Nevertheless, vortex entry and edge-barrier effects may still contribute to the switching instability. To evaluate whether magnetic flux can penetrate the nanoring, we estimated the characteristic field $B_s$ for stable vortex entry in the 118-nm-wide nanowire using[41] :

$$B_s = \frac{2\phi_0}{\pi W^2} \ln\left[\frac{\alpha W}{\xi(T)}\right] \quad (1)$$

where $W = 118$ nm, $\alpha = 0.1043$, and $\xi(2\text{ K}) \approx 4.5$ nm, yielding $B_s \approx 951$ G. This confirms that, in the field range applied in our experiments (up to 2 kG), magnetic flux quanta can indeed enter the nanoring. The absence of observable flux-flow resistance in the region $0 < I < I_c$ is therefore not due to insufficient flux entry; rather, it may be attributed to the narrow nanowire width, which creates a strong edge barrier. This results in a large vortex depinning current: once vortices begin to move under higher driving current, the associated Joule heating rapidly drives the entire nanowire into the normal-resistance state before a clear flux-flow regime can develop. The above results suggests that the observed nonreciprocal critical currents likely arise from an interplay among curvature-induced current crowding, Meissner screening, asymmetric vortex-entry barriers, and abrupt superconducting switching.

To explicitly demonstrate the essential role of the engineered geometric asymmetry, we have now fabricated a control device with a symmetric nanowire. As shown in Fig. 1(d), at $B = 400$ G, the $I$-$V$ characteristics of this symmetric device show only negligible nonreciprocity in the critical currents.

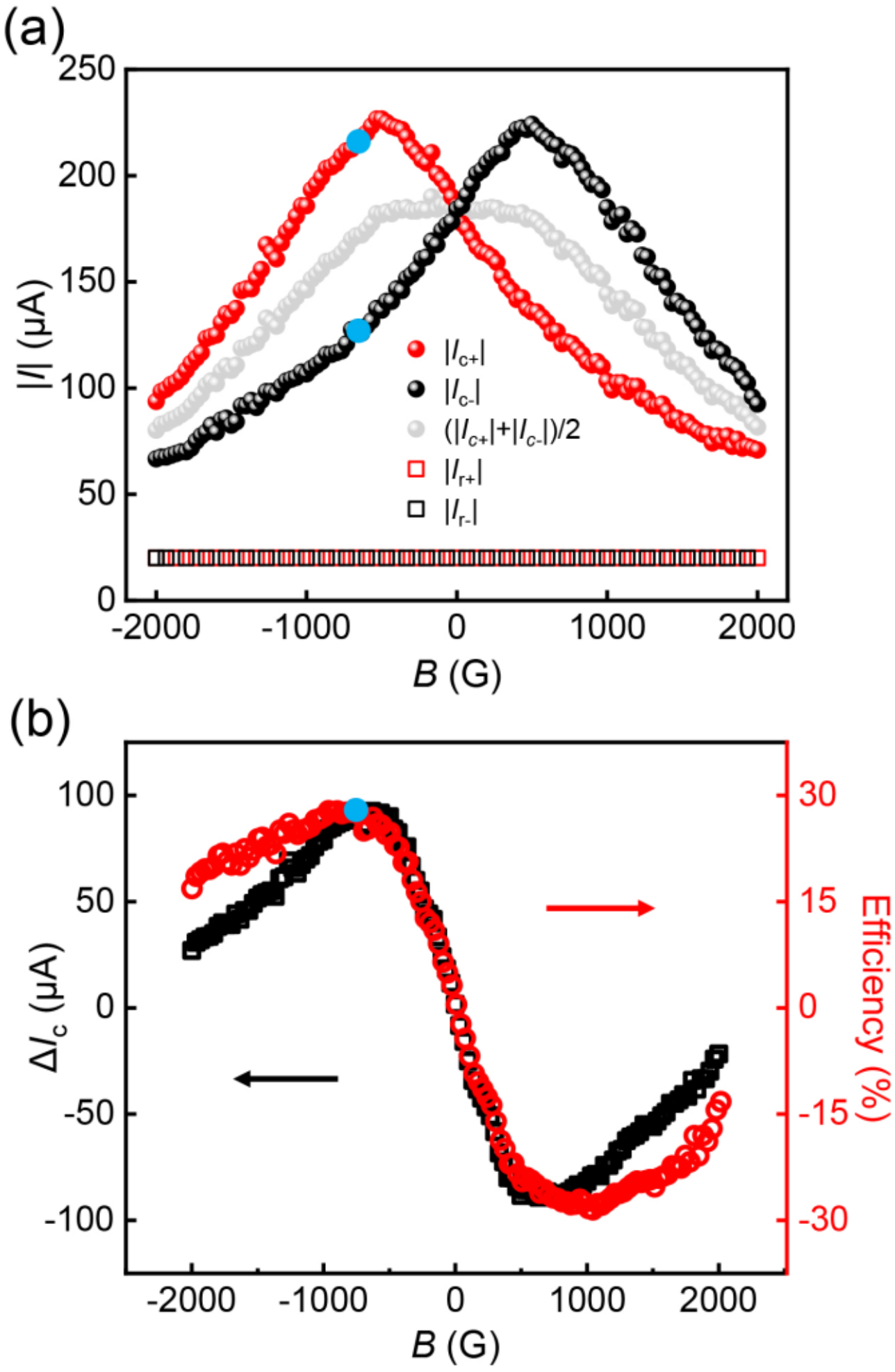

**Figure 2. Magnetic field dependence of nonreciprocity.** (a) Magnetic-field dependence of the critical current ($I_c$) and retrapping current ($I_r$) measured for opposite current directions at $T = 2$ K. The gray dots represent the average of the positive and negative critical current magnitudes. (b) Nonreciprocal critical current ($\Delta I_c$) and diode efficiency ($\eta$) as a function of the magnetic field.

In addition to the nonreciprocal critical currents, we also examined the retrapping current $I_r$ extracted during the normal-to-superconducting transition when sweeping down the current [Figs. 1(b)-1(d)]. Unlike the critical current, the retrapping current $I_r = 20$ μA remains essentially identical for both current polarities, showing no nonreciprocity and being insensitive to the applied magnetic field [Fig. 2(a)]. To rule out the possibility that the film behaves as a disorder-induced Josephson-junction array, which can occur in granular films with nanometer thickness, we estimated the critical current density from our devices. The extracted value, $J_c \approx 1.7 \times 10^{11}$ A/m², is consistent with that reported for high-quality, uniform NbN films[38] and is far higher than the typical $J_c$ values of granular NbN films or Josephson-junction arrays (on the order of $10^7$ A/m²)[39]. The observed hysteresis is therefore unlikely to originate from such granular behavior.

The observed hysteretic behavior is consistent with a thermally dominated retrapping process [38]. The perfect symmetry of $I_{c+}$ and $I_{c-}$ at zero applied field [Fig. 2(a)] confirms that remanent magnetic fields from the superconducting magnet are negligible and do not affect the observed field-independent retrapping current. While the critical current is determined by the onset of superconductivity breakdown—strongly influenced by geometric current distribution, vortex entry, and symmetry breaking—the retrapping current is likely governed by thermal dynamics in the resistive state[42]. After the device switches to the normal state, Joule heating raises the local electron temperature, and the return to the superconducting state occurs only after the device cools sufficiently below the equilibrium $T_c$. The minimal Joule heating ($P = I_r^2 R_n \approx 0.52$ μW, given the normal state resistance $R_n \approx 1.3$ kΩ) that holds the nanoring at a local temperature above $T_c$ determines the retrapping current $I_r$. Consequently, $I_r$ is primarily set by thermal relaxation and heat diffusion processes within the NbN film and the substrate, which are largely independent of magnetic field and insensitive to geometric asymmetry. This explains why $I_r$ in a superconducting diode remains reciprocal and nearly field-independent, even though the critical current exhibits strong nonreciprocity. The coexistence of nonreciprocal $I_c$ and reciprocal $I_r$ is therefore consistent with a thermally dominated retrapping mechanism that is decoupled from the symmetry-breaking processes responsible for the diode effect.

To investigate the magnetic-field dependence of the diode effect, we performed systematic transport measurements at $T = 2$ K while sweeping the perpendicular magnetic fields. The extracted critical currents $I_{c+}$ and $I_{c-}$ are plotted as a function of magnetic fields in Fig. 2(a). Under finite fields, two branches exhibit clear nonreciprocity, in contrast to their symmetric behavior at zero field. Notably, at low magnetic fields ($|B| \lesssim 600$ G), the average critical current $(|I_{c+}| + |I_{c-}|)/2$ remains approximately constant, as indicated by the gray dots in Fig. 2(a), even though $I_{c+}$ and $I_{c-}$ shift in opposite directions with increasing magnetic fields. This behavior indicates that the magnetic field redistributes the current-carrying capability asymmetrically between the two bias directions—enhancing one branch while suppressing the other—without significantly reducing the overall superconducting strength. This redistribution can be understood through a phenomenological model of current superposition with Meissner screening current under small magnetic field[6]. In our π-shaped nanoring, geometric crowding naturally concentrates the transport current at the inner edge, where it superimposes with the magnetic-field-induced Meissner screening current $I_s(B)$. Because this screening current adds to the local transport current in one bias direction and subtracts in the other, it linearly shifts the observed critical currents, $I_{c\pm} \approx \pm I_{c0} - \delta I_s(B)$, while keeping the average constant. Such conservation of the average critical currents suggests that the magnetic field primarily induces a directional modulation of the superconducting switching instability, rather than uniform depairing or degradation of superconductivity, providing a natural origin for the emergence of critical-current nonreciprocity in the low-field regime.

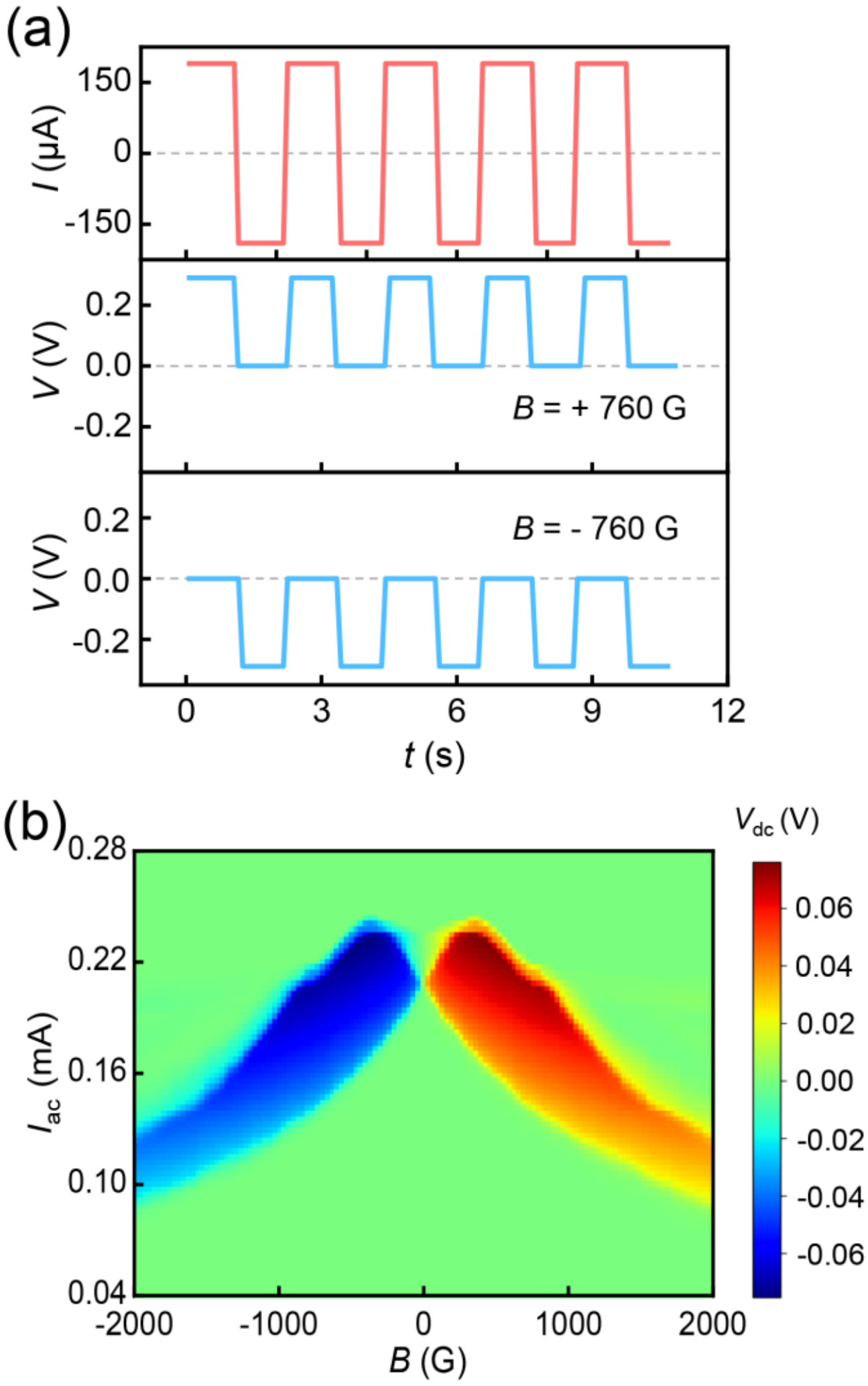

**Fig. 3. Rectification effect.** (a) Half-wave rectification measured under a square-wave current drive (top panel). Measurements were performed at 2 K under magnetic fields of +760 G (middle panel) and -760 G (bottom panel). The magnetic field of 760 G (Corresponding to the blue dots

highlighted in Figure 2) was chosen as it yields the highest diode efficiency at 2 K. (b) Colormap showing the DC rectification voltage as a function of the sinusoidal alternating current ($f$ = 5353 Hz) and the magnetic field.

One important application of a superconducting diode is signal rectification. To directly evaluate this functionality, we applied a square-wave current excitation with an amplitude of 190 μA, which is between the $|I_{c+}|$ and $|I_{c-}|$ [top panel of Fig. 3(a)]. The device switches reproducibly between the superconducting and normal states [the middle panel of Fig. 3(a)], and the switching polarity reverses upon reversing the applied magnetic field [bottom panel of Fig. 3(a)]. This demonstrates controllable unidirectional supercurrent transport consistent with the field-dependent diode effect measured in Fig. 2. We further assessed the rectification response by applying a sinusoidal AC current (5353 Hz). The resulting DC output voltage, mapped in Fig. 3(b), confirms efficient rectification over a broad range of magnetic fields.

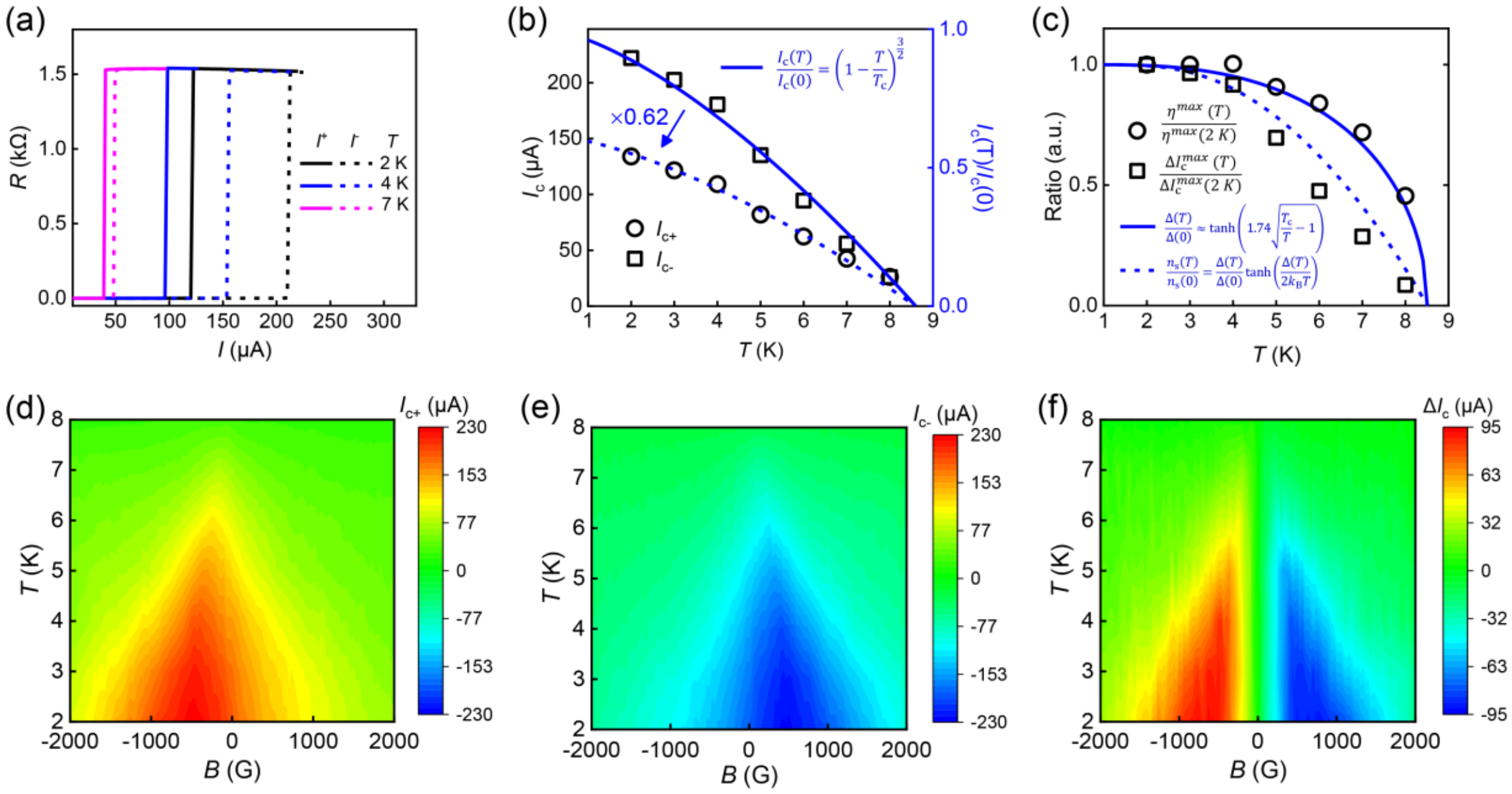

**Figure 4. Temperature dependence of nonreciprocity.** (a) Current-voltage curves for positive (solid lines) and negative (dashed lines) bias measured at various temperatures in a magnetic field of 700 G. (b) Temperature dependence of the critical current obtained under positive (black circles) and negative (black squares) bias measured at a magnetic field of 500 G. The solid blue curve is a fit based on the theoretical Ginzburg-Landau depairing critical current formula. The dashed blue line is the solid blue curve shifted by a ratio of 0.62, representing the same temperature evolution for the positive and negative critical currents. (c) Maximal normalized nonreciprocal diode efficiencies (black circles) and critical currents (black squares) extracted from (d-f) at various temperatures. The solid blue curve represents the theoretical calculation of the temperature dependence of the BCS superconducting gap, and the dashed blue curve shows the calculated temperature dependence of the superfluid density. (d)-(f) Colormaps of the positive critical current $I_{c+}$ (d), negative critical current $I_{c-}$ (e) and nonreciprocal critical current $\Delta I_c$ (f) as a function of temperature and magnetic field.

The magnitude of the nonreciprocity, defined as $\Delta I_c = |I_{c+}| - |I_{c-}|$, is displayed by the black curve in Fig. 2(b). $\Delta I_c$ s increase with increasing field magnitudes, reach a maximum, and then decrease, forming an antisymmetric profile satisfying $\Delta I_c(B) = -\Delta I_c(-B)$. This antisymmetry reflects the reversal of diode polarity with the reversal of the magnetic-field direction, consistent with the behavior of the individual critical-current branches. The diode efficiency, defined as[3, 6, 9-12, 14, 16, 17]

$$\eta = \frac{|I_{c+}| - |I_{c-}|}{|I_{c+}| + |I_{c-}|} \quad (2)$$

is shown by the red curve in Fig. 2(b). The efficiency follows a similar field dependence and also changes sign with magnetic-field reversal. At 2 K, the device reaches a maximum diode efficiency of 28.1%, demonstrating strong nonreciprocal transport in a simple single-layer NbN nanoring structure.

Furthermore, the rectification effect was systematically verified at multiple temperatures from 2 K to 7 K below $T_c$, where robust rectification persists across the entire range (Fig. S2). Device-to-device reproducibility was confirmed on three independent nanoring devices, and the rectification remains stable over at least 100 cycles (Figures S3 and S4). While these results illustrate the potential of the geometrically asymmetric NbN nanoring as a simple and lithography-compatible platform, realizing large-scale integration in cryogenic superconducting electronics will require further detailed studies on the frequency dependence of rectification, additional performance optimization, and strategies for zero-field operation (e.g., via hybrid integration with magnetic materials).

To examine the temperature dependence of the diode effect, we measured the diode device over a range of temperatures below $T_c$. The *I-R* characteristics retain sharp switching behavior at all temperatures [Fig. 4(a)]. The field-dependent critical currents — $I_{c+}$ and $I_{c-}$ — measured at various temperatures ranging from 2 K to 8 K are shown in Figs. 4(d) and 4(e), respectively. Both $|I_{c+}|$ and $|I_{c-}|$ decrease monotonically with increasing temperature [Fig. 4(b)], consistent with thermal suppression of superconductivity. As shown in Fig. 4(b), their temperature dependence follows the Ginzburg-Landau depairing critical current relation[43] as an empirical form,

$$I_c(T) = I_c(T = 0)\left(1 - \frac{T}{T_c}\right)^{\frac{3}{2}} \quad (3)$$

While this empirical agreement suggests that a nonreciprocal depairing process is a plausible underlying mechanism for the observed diode effect, we note that in our specific geometry — a narrow asymmetric nanoring under a perpendicular magnetic field—complex vortex dynamics cannot be ruled out. Specifically, asymmetric vortex entry barriers, fluxoid states, and geometric current crowding likely play relevant roles [44]. Therefore, the observed nonreciprocity may represent an interplay between asymmetric depairing and vortex-ratchet contributions.

To further explore the relationship between superconducting parameters and the observed diode effect, we plot the temperature-field maps of the nonreciprocity $\Delta I_c$ in Fig. 4(f). The corresponding maximal values, $\Delta I_c^{max}$ and the extracted $\eta^{max}$, obtained at various temperatures are shown in Fig. 4(c). Although the $\Delta I_c^{max}(T)$ and $\eta^{max}(T)$ exhibit distinct temperature dependence, they both decrease monotonically with increasing temperature, indicating that the overall superconducting diode effect is closely linked to the current-carrying capability of the device.

As the depairing critical current of a given superconductor is determined by both the Cooper-pair binding energy—characterized by the superconducting energy gap $\Delta(T)$—and the superfluid density $n_s(T)$, we compare the experimentally extracted $\Delta I_c^{max}(T)$ and $\eta^{max}(T)$ with the theoretical temperature dependence of the BCS energy gap[45] and superfluid density[46], given by

$$\frac{\Delta(T)}{\Delta_0} \approx \tanh\left(1.74\sqrt{\frac{T_c}{T} - 1}\right), \quad (3)$$

$$\frac{n_s(T)}{n_s(0)} = \frac{\Delta(T)}{\Delta_0}\tanh\left(\frac{\Delta(T)}{2k_B T}\right). \quad (4)$$

As shown in Fig. 4(c), we observe a phenomenological correlation where $\eta^{max}(T)$ closely tracks the temperature evolution of the superconducting energy gap $\Delta(T)$, while $\Delta I_c^{max}(T)$ exhibits a temperature dependence more similar to the superfluid density $n_s(T)$. While these correlations suggest that the magnitude of critical-current nonreciprocity and the diode efficiency may be influenced by different superconducting parameters, further studies are needed to determine if these metrics are governed independently by these parameters. Nevertheless, this comparison provides suggestive guidance for material selection in the design of superconducting diodes tailored to specific application requirements.

## 3. Conclusion

In conclusion, we have demonstrated that a pronounced geometrically induced superconducting diode effect in the structurally minimal, single-material NbN nanoring. Distinct from superconducting diode devices based on Josephson junctions, multilayers, ferromagnetic hybrids, gate-controlled structures, or artificial pinning arrays, the present device realizes nonreciprocal critical-current switching through the asymmetric annular geometry of the superconducting nanowire itself. Under a perpendicular magnetic field, the device exhibits pronounced and polarity-switchable critical-current nonreciprocity, while maintaining symmetric and field-independent retrapping currents. This behavior indicates that the superconducting diode effect originates from a directional modulation of the superconducting switching instability rather than from thermal effects or uniform suppression of superconductivity. Together with the negligible nonreciprocity observed in the symmetric control device, these results confirm that curvature-induced edge asymmetry provides the essential inversion-symmetry breaking responsible for the diode effect.

Systematic magnetic-field- and temperature-dependent measurements further reveal that, at low magnetic fields, the applied field redistributes the critical current asymmetrically between opposite bias directions without significantly reducing the overall superconducting current-carrying capability. Importantly, while both the maximal critical-current nonreciprocity $\Delta I_c^{max}$ and the maximal diode efficiency $\eta^{max}$ decrease monotonically with increasing temperature, they exhibit distinct temperature dependences: the $\eta^{max}$ follows the evolution of the superconducting energy gap, whereas the $\Delta I_c^{max}$ is more closely associated with the superfluid density.

In addition to its clear rectification functionality, the compact and fully planar nanoring architecture is compatible with standard lithographic processes. Together, these findings establish asymmetric superconducting nanorings as a simple and effective minimal geometric platform for studying nonreciprocal superconducting transport, and suggest clear routes for device optimization through both material selection and geometric design, depending on the specific performance requirements of targeted applications. While practical circuit-level implementation will require further optimization toward lower-field or zero-field operation, the present work provides a clear geometric design principle for future superconducting rectifying elements.

## 4. Method

**Device fabrication.**

The superconducting nanoring devices were fabricated from a 10-nm-thick NbN film deposited on a commercial Si/$SiO_2$ substrate by magnetron sputtering. The NbN film was first patterned by ultraviolet lithography and followed by reactive-ion etching to define the coarse device geometry. Ti/Au electrodes were then deposited to form low-resistance electrical contacts. The asymmetric nanoring, with an inner diameter of 600 nm and a linewidth of approximately 118 nm, was defined by electron-beam lithography followed by reactive-ion etching. The device morphology was characterized by field-emission scanning electron microscopy. Detailed fabrication parameters, including sputtering, lithography, and etching conditions, are provided in the Supporting Information.

**Transport measurements.**

Electrical transport measurements were performed using a standard four-probe configuration in a superconducting cryogenic magnet system. The magnetic field was applied perpendicular to the sample plane. A Keithley 6221 current source was used to generate DC, quasi-DC, square-wave, and sinusoidal current excitations, while the voltage response was measured using a Keithley 2182A nanovoltmeter. Measurements were performed at fixed temperatures with a temperature stability better than ±1 mK.

**Data analysis.**

The positive and negative critical currents, $\boldsymbol{I}_{\mathbf{c}+}$ and $\boldsymbol{I}_{\mathbf{c}-}$, were extracted from the current–voltage characteristics at the superconducting-to-resistive switching points. The retrapping currents, $\boldsymbol{I}_{\mathbf{r}+}$ and $\boldsymbol{I}_{\mathbf{r}-}$, were extracted from the resistive-to-superconducting transition during the reverse current sweep. The nonreciprocal critical-current component was defined as $\Delta\boldsymbol{I}_{\mathbf{c}} = |\boldsymbol{I}_{\mathbf{c}+}| - |\boldsymbol{I}_{\mathbf{c}-}|$, and the diode efficiency was calculated as $\boldsymbol{\eta} = [|\boldsymbol{I}_{\mathbf{c}+}| - |\boldsymbol{I}_{\mathbf{c}-}|]/[|\boldsymbol{I}_{\mathbf{c}+}| + |\boldsymbol{I}_{\mathbf{c}-}|]$.

## Acknowledgments

This work was supported by the Quantum Science and Technology-National Science and Technology Major Project (Grant 2024ZD0301300), the National Key R&D Program of China (Grants 2021YFA0718802 and 2024YFA1408900), the National Natural Science Foundation of China (Grants 62274086, 62288101, 62571230, 92565201, and 12504137), the China Postdoctoral Science Foundation (Grants 2024M751370, 2025T180930), the Jiangsu Funding Program for Excellent Postdoctoral Talent (Grant 2023ZB534), the CAS Superconducting Research Project (Grant SCZX-0201), and the Natural Science Foundation of Jiangsu Province (Grant BK20241223).